\documentclass[amsmath,amssymb,twocolumn]{revtex4}
\usepackage{dcolumn}
\usepackage{bm}
\usepackage{amsmath,mathrsfs}
\usepackage{graphicx,epsfig}
\usepackage{hyperref}
\usepackage{epsf}
\usepackage{color}

\begin{document}


\title{{\bf Wormholes in $f(R,T)=R+\lambda T+\lambda_1 T^2$ gravity}}

\author{ F. Parsaei }\email{fparsaei@gmail.com}
\author{S. Rastgoo}\email{rastgoo@sirjantech.ac.ir}

\affiliation{ Physics Department , Sirjan University of Technology, Sirjan 78137, Iran.}

\date{\today}


\begin{abstract}
\par   his study explores asymptotically flat wormhole solutions within the framework of $f(R,T)$ gravity. We analyze $f(R,T)$ expressed as $f(R,T)=R+\lambda T+\lambda_1 T^2$. A linear equation of state is employed for both radial and lateral pressures, resulting in a power-law shape function. The investigation encompasses solutions characterized by both negative and positive energy densities. It has been determined that solutions with positive energy density comply with all energy conditions, specifically the null, weak, strong, and dominant energy conditions. Additionally, we identify constraints on the parameters $\lambda$, $\lambda_1$, and the parameters associated with the equation of state and shape function. \\
\end{abstract}

\maketitle
\section{Introduction}

Wormholes are solutions to Einstein's field equations that connect two universes or two distant regions within the same universe. The initial wormhole solutions examined by Flamm, were unstable, making it impossible for observers to traverse them \cite{flamm}. Einstein and Rosen mathematically described the structure of the Einstein-Rosen bridge; however, it is not a traversable wormhole \cite{Rosen}.  Misner and Wheeler introduced the term “wormhole” in 1957 \cite{wheeler}. Wormholes might be sufficiently large for humanoid travelers and could even facilitate time travel \cite{WH}. A traversable wormhole solution is devoid of any horizon or singularity. Ellis  uncovered a novel wormhole solution for a spherically symmetric configuration of Einstein’s equations, incorporating a massless scalar field with ghost-like properties \cite{Ellis}. A crucial aspect of wormhole formation within the framework of general relativity (GR) is the transgression of energy conditions \cite{Visser}. Such violations of energy conditions frequently encounter skepticism, contributing to a widely held belief, at least historically, that wormholes are not physically feasible entities. Consequently, a matter distribution that contravenes the null energy condition (NEC) is classified as exotic and is considered to possess minimal physical significance, primarily due to the lack of experimental evidence supporting its existence.

 One of the most intriguing findings in contemporary cosmology is the accelerated expansion of the Universe. This phenomenon has garnered significant attention from researchers in the field. Following this discovery, extensive investigations into phantom wormholes that contravene energy conditions (ECs) have been conducted in the literature \cite{phantom, phantom2, phantom1}. Addressing the challenge of minimizing the reliance on exotic matter in the construction of wormholes is another critical aspect of this research. Scholars are endeavoring to restrict the use of exotic matter to a limited region of spacetime. Examples of this include thin shell wormholes \cite{cut, cut1, cut2, cut3}, wormholes characterized by a variable equation of state (EoS) \cite{Remo, variable}, and those defined by a polynomial EoS \cite{foad}.

Recently, an alternative approach to wormhole physics has been investigated within the framework of modified gravity. The extra components of the gravitational part maintain the geometry of the traversable wormhole  while ensuring that the matter components remain non-exotic. In this realm, wormholes have been studied extensively in modified theories. Wormholes are studied in  Braneworld \cite{b, b1, b2, b3}, Born-Infeld theory \cite{Bo, Bo1}, quadratic gravity \cite{quad, quad1}, Einstein-Cartan gravity \cite{Cartan, Cartan1, Cartan2}, Rastall–Rainbow gravity \cite{RaR, RaR1},  $f(Q)$ gravity \cite{fq, fq1, fq2, fq3, fq4, fq44, fq5, fq6} , $f(R)$ gravity \cite{Nojiri, fR0, fR11, fR22, fR33, fR44,fR55} and Ricci inverse gravity \cite{inverse}. Some of these modified theories can resolve the problem of exotic matter for  wormholes. The common foundation of most modified theories of gravity is that they are Lagrangian theories, meaning that all these proposals are formulated through various generalizations of the Einstein-Hilbert action.

In this work, we are interested in a covariant generalization of Einstein’s GR known as $f(R,T)$ gravity. This theory is an extension of $f(R)$ gravity which the standard Einstein-Hilbert action is replaced by an arbitrary function of the Ricci scalar ($R$) \cite{fRR}. Coupling any function of the Ricci scalar $R$ with the matter Lagrangian density  $Lm$ generates  $f (R, T)$ theory of gravity \cite{RT, RTcoment}.

 The cosmological phenomena within the framework of $f(R,T)$ have been examined in the literature \cite{cosm, cosm1, cosm2}. Black holes have been explored  in the framework of $f(R, T)$ gravity \cite{Black, Black2}. Numerous studies focus on wormhole solutions in the context of $f(R,T)$ gravity. Different forms of $f(R,T)$ functions have been considered by researchers to investigate wormhole solutions. Azizi \cite{Azizi} has derived a shape function based on the assumption of a linear equation of state for matter and presents solutions that fulfill the energy conditions. The modeling of wormholes in $f(R, T)$ gravity is discussed in \cite{Moa} by Moraes and Sahoo. They have discerned solutions characterized by a linear EoS and those with a variable EoS parameter. In their study, the solutions exhibiting a linear EoS adhere to the ECs, while those with a variable EoS fail to meet these conditions. Zubair et al. investigate asymptotically flat wormhole solutions within the framework of the $f(R,T)$ modified theory of gravity, employing established non-commutative geometry through Gaussian and Lorentzian distributions that are derived from string theory \cite{Zub}. Sharif and Nawazish explored wormhole solutions within spherically symmetric spacetime utilizing the Noether symmetry method in the framework of $f(R, T)$ gravity \cite{Shar}. In \cite{Shw}, the auothers employed two different traversable wormhole geometries characterized by exponential and power-law shape functions for modeling the wormholes. Additionally, in \cite{Cha}, three separate models of $f(R,T)$ are analyzed to uncover precise wormhole solutions. The topic of charged wormhole solutions within the framework of the $f(R,T)$ extended theory of gravity has been explored in \cite{Charge}. Additionally, \cite{Sharif} examines traversable wormhole solutions that adhere to the Karmarkar condition in the context of $f(R,T)$ theory. Furthermore, Rosa and Kull have demonstrated that traversable wormhole solutions, characterized by a non-vanishing redshift function in the linear formulation of $f(R,T)=R + \lambda T$ gravity, satisfy the energy conditions throughout the entire spacetime \cite{Rosa}. A novel hybrid shape function for a wormhole within the framework of modified $f(R, T)$ gravity has been introduced in \cite{Sah}. Tripathy et al. have examined a model with $f(R,T)=R+\lambda T+ \lambda_1 T^2$  featuring power-law and exponential shape functions \cite{squared}. They demonstrated that the existence of non-exotic matter traversable wormholes is not evident in the model, and its feasibility may hinge on the selection of the wormhole geometry. Additionally, they discovered that non-exotic wormholes are achievable within the specified squared trace extended gravity theory for a limited range of the chosen EoS parameter. Numerous other wormhole solutions have been explored within the framework of $f(R,T)$ gravity \cite{Moa2, Zub2, fr1, fr2, fr3, fr4, Yousaf, Sha, Sahoo, god, Bha, Ban, Mish, Tang, Noori, Sama, Gosh, Sarkar, Chau}. Recently wormholes in a generalized geometry-matter coupling theory of gravity, $f(R,L,T)$ are investigated \cite{RLT, RLT1}.

 In the current body of literature, solutions characterized by a constant redshift function that comply with the energy conditions are primarily restricted to a power-law shape function and a linear EoS for the model $f(R, T)=R+\lambda T$ \cite{Azizi, Moa}, and recently variable EoS \cite{SR}. This study aims to explore wormhole structures across a spectrum of nonlinear functional forms represented by $f(R, T)=R+\lambda T+\lambda_1 T^2$. The breadth of this investigation is considerably more extensive than the wormhole analysis previously conducted by Tripathy et al. \cite{squared}.

The structure of this document is organized as follows: Section \ref{sec2} explores the criteria and equations that define wormholes. Following this, we present a brief overview of $f(R, T)$ theory alongside the classical ECs. In Section \ref{sec3}, we utilize the field equations to derive the shape function within the framework of $f(R, T)$ gravity, presenting solutions that satisfy the energy conditions. This section also includes an examination of the physical properties associated with these solutions. Finally, we conclude with our remarks in the last section. Throughout this paper, we operate under the assumption of gravitational units, specifically $c = 8 \pi G = 1$.

\section{Basic formulation of wormhole and $f(R,T)$ gravity} \label{sec2}
The line element of a static and spherically symmetric metric can be written as
\begin{equation}\label{1}
ds^2=-U(r)dt^2+\frac{dr^2}{1-\frac{b(r)}{r}}+r^2(d\theta^2+\sin^2\theta,
d\phi^2)
\end{equation}
where $U(r)=\exp (2\phi(r))$.
The metric function $b(r)$ is called the shape function that determines the shape of the wormhole. Here, $ \phi(r)$ is called the redshift function which can be used to detect the redshift of the signal  by a distance observer.
The condition
\begin{equation}\label{2}
b(r_0)=r_0.
\end{equation}
where  $r_0$ is the wormhole throat must be held at the wormhole throat.  Additionally, two additional conditions must be satisfied to ensure the existence of a traversable wormhole,
\begin{equation}\label{3}
b'(r_0)<1
\end{equation}
and
\begin{equation}\label{4}
b(r)<r,\ \ {\rm for} \ \ r>r_0.
\end{equation}
Equation (\ref{3}) leads to the violation of NEC in the background of GR which is well-known as the flaring-out condition.
In this work, we aim to obtain asymptotically flat geometries. Therefore, the metric functions must satisfy the following conditions:
\begin{equation}\label{5}
\lim_{r\rightarrow \infty}\frac{b(r)}{r}=0,\qquad   \lim_{r\rightarrow \infty}U(r)=1.
\end{equation}
For simplicity, we have concentrated on solutions featuring a constant redshift function. In this article, we examine an anisotropic fluid in the form  of  $T^{\mu}_{\nu}=diag[-\rho, p_r,p_t,p_t]$, where $\rho$ denotes the energy density, $p_r$ is the
 radial pressure and $p_t$ denotes the tangential pressure, respectively.

Let us take a moment to briefly review the $f(R, T)$ formalism. The field equation corresponding to the $f(R, T)$ gravity model is derived from the Hilbert-Einstein action as follows:
\begin{equation}\label{6}
S=\int\frac{1}{2}f(R,T)\sqrt{-g}\;d^{4}x+\int L_{m}\sqrt{-g}\;d^{4}x.
\end{equation}
where $f(R, T)$ is a general function  of $R$ (Ricci scalar) and $T$ (trace of the energy-momentum tensor), $g$ is the determinant
of the metric, and $L_m$ is the matter Lagrangian density. This geometrically modified action features a well-behaved function of $f(R,T)$ that replaces the conventional Ricci scalar $R$ in the Einstein-Hilbert action. The relationship between $L_m$ and the energy-momentum tensor is expressed by
\begin{equation}\label{7}
T_{ij}= - \frac{2}{\sqrt{-g}}\left[\frac{\partial(\sqrt{-g}L_{m})}{\partial g^{ij}}-\frac{\partial}{\partial x^{k}}\frac{\partial(\sqrt{-g}L_m)}{\partial(\partial g^{ij}/\partial x^{k})}\right].
\end{equation}
We assume $L_m$ is dependent only on the metric component not on its derivatives so
\begin{equation}\label{8}
T_{ij}= g_{ij}L_{m} - 2\frac{\partial L_{m}}{\partial g^{ij}}.
\end{equation}
A variation of the modified action with respect to the metric gives
\begin{multline}\label{9}
f_R(R,T)\left(R_{ij}-\frac{1}{3} Rg_{ij}\right) + \frac{1}{6}f(R,T)g_{ij} \\= \left(T_{ij}-\frac{1}{3}Tg_{ij}\right)-f_T(R,T)\left(T_{ij} -\frac{1}{3}Tg_{ij}\right)\\-f_T(R,T)\left(\theta_{ij}-\frac{1}{3}\theta g_{ij}\right)+\nabla_i\nabla_jf_R(R,T),
\end{multline}
 where $f_R (R,T)\equiv \frac{\partial f(R,T)}{\partial R}$, $f_T (R,T)\equiv \frac{\partial f(R,T)}{\partial T}$ and
 \begin{equation}\label{10}
\theta_{ij}=g^{ij}\frac{\partial T_{ij}}{\partial g^{ij}}.
\end{equation}
$L_m = -T$, $L_m =P$ where $P=\frac{p_r+2p_t}{3}$, and $L_m =-\rho$  are options for matter lagrangian density with different physical interpretation. We assume $L_m =P$ which is a natural choice.  Hence, Eq.(\ref{10}) gives
\begin{equation}\label{11}
\theta_{ij}=-2T_{ij}+P g_{ij}.
\end{equation}
The functional $f(R, T)$ can be represented as the sum of two distinct functions, and we may have
\begin{equation}\label{13}
f(R,T)=f_1(R)+f_2(T),
\end{equation}
or
\begin{equation}\label{13a}
f(R,T)=f_1(R)+f_2(R)f_3(T).
\end{equation}
The most straightforward scenario is as follows
\begin{equation}\label{14}
f(R,T)=R+\lambda T,
\end{equation}
which is studied in the literature extensively \cite{Azizi, Moa, Zub, Rosa, SR}. Generally, the $f(R,T)$ in the form of (\ref{14}) leads to a  change in effective Einstein's gravitational constant  which can resolve the problem of exotic matter for wormhole theory \cite{ SR}. In this article, we are intersected to find solutions in the context of
 \begin{equation}\label{14a}
f(R,T)=R+\lambda T+\lambda_1 T^2,
\end{equation}
which is well-known as squared trace gravity theory. In this case, by using (\ref{1}), (\ref{9}), (\ref{11}) and (\ref{14a}), one can find the following field equations
\begin{eqnarray}
\frac{b^{\prime}(r)}{r^2}&=& \rho+\frac{2\lambda}{6}[9\rho-(p_r+2p_t)]+\nonumber\\
\frac{\lambda_1}{6}&[&(\rho-p_r-2p_t)(15\rho+p_r+2p_t)],\label{15}\\
\frac{b(r)}{r^3}&=&- p_r+\frac{\lambda}{6}\left[3\rho-(7p_r+2p_t)\right]+\nonumber\\
\frac{\lambda_1}{6}&[&(\rho-p_r-2p_t)(3\rho-11p_r+2p_t)],\label{16}\\
\frac{b^{\prime}(r)}{r^2}&-&\frac{b(r)}{r^3} = -2 p_t+\frac{\lambda}{3}\left[3\rho-(p_r+8p_t)\right]+\nonumber\\
\frac{\lambda_1}{3}&[&(\rho-p_r-2p_t)(3\rho+p_r-10p_t)].\label{17}
\end{eqnarray}
which the prime denotes the derivative $\frac{d}{dr}$. Now, we have three equations and four unknown functions, $\rho(r),\, p_r(r),\, p_t(r)$ and $b(r)$. It is important to note that the field equations are generally non-linear, so introducing an additional equation may not resolve the issue. Conversely, arbitrarily considering an unknown function may not yield consistent solutions. We consider an EoS in the following form
\begin{equation}\label{24}
p_r (r)=w_r\rho
\end{equation}
and
\begin{equation}\label{241}
p_t (r)=w_t\rho.
\end{equation}
This choice for EoS is one of the most favored among physicists. In the next section, we examine the possible solutions for this EoS.

Wormhole solutions in the context of standard General Relativity have been demonstrated to contravene energy conditions. To ensure a positive stress-energy tensor when matter is present, these energy conditions provide effective approaches. The energy conditions, which include the NEC, Dominant Energy Condition (DEC), Weak Energy Condition (WEC), and Strong Energy Condition (SEC), are clearly articulated to facilitate the attainment of this objective.
\begin{eqnarray}\label{21}
\textbf{NEC}&:& \rho+p\geq 0,\quad \rho+p_t\geq 0 \\
\label{21a}
\textbf{WEC}&:& \rho\geq 0, \rho+p\geq 0,\quad \rho+p_t\geq 0, \\
\textbf{DEC}&:& \rho\geq 0, \rho-|p|\geq 0,\quad \rho-|p_t|\geq 0, \\
\textbf{SEC}&:& \rho+p\geq 0,\, \rho+p_t\geq 0,\rho+p+2p_t \geq 0. \label{21b}
\end{eqnarray}
 Derived from the Raychaudhuri equations, these conditions are essential tools for comprehending the geodesics of the Universe. By employing these equations, we can decipher the intricate paths taken by cosmic objects. Now, as stated in \cite{fq} by defining the functions,
\begin{eqnarray}\label{22}
 H(r)&=& \rho+p ,\, H_1(r)= \rho+p_t,\, H_2(r)= \rho-|p|, \nonumber \\
 H_3(r)&=&\rho-|p_t|,\, H_4(r)= \rho+p+2p_t ,
\end{eqnarray}
we can examine the ECs in the later sections of this paper. For the sake of simplicity, we will assume that $r_0=1$ in the subsequent parts of this document.

\section{Wormhole solutions }\label{sec3}

 In this section, we explore the potential wormhole solutions with a linear EoS and $f(R,T)$ as expressed in (\ref{14a}). The procedure is similar to reference \cite{squared}.
  Using (\ref{24}) and (\ref{241}) in (\ref{15}-\ref{17}) gives
 \begin{eqnarray}
\frac{b^{\prime}(r)}{r^2} &=& \alpha_1\rho+\beta~\alpha_2\rho^2,\label{18}\\
\frac{b(r)}{r^3} &=& \alpha_3\rho+\beta~\alpha_4\rho^2,\label{19}\\
\frac{b^{\prime}(r)}{r^2}-\frac{b(r)}{r^3} &=& \alpha_5\rho+\beta~\alpha_6\rho^2,\label{20}
\end{eqnarray}
where
\begin{eqnarray}
\alpha_1 &=&  \frac{\lambda}{6}\left(9-\omega_r-2\omega_t\right)+1 , ~\alpha_2 = \omega_r+2\omega_t+15, \nonumber\\
\alpha_3 &=& \frac{\lambda}{6}\left(3-7\omega_r-2\omega_t\right)-\omega_r, ~\alpha_4 = 3-11\omega_r+2\omega_t, \nonumber \\
\alpha_5 &=& \frac{\lambda}{3} \left(3-\omega_r-8\omega_t\right)-2\omega_t, ~\alpha_6 = 2\left(3+\omega_r-10\omega_t\right),\nonumber\\
\beta &=& \frac{\lambda_1}{6} \left(1-\omega_r-2\omega_t\right).  \label{20a}
\end{eqnarray}
Equations. (\ref{18}-\ref{20}) are consistent while
\begin{eqnarray}
\alpha_1-\alpha_3 &=& \alpha_5,\label{20b}\\
\beta (\alpha_2-\alpha_4) &=& \beta \alpha_6.\label{20c}
\end{eqnarray}
Equation (\ref{20b}) yields
\begin{equation}\label{24a}
\omega_r=-(2\omega_t +\frac{6}{8\lambda +6}).
\end{equation}
Equation (\ref{20c}) for  $\beta\neq0$ gives
\begin{equation}\label{24b}
\omega_t=-\frac{\omega_r}{2}-0.3.
\end{equation}
It is easy to show that Eqs.(\ref{24a}) and (\ref{24b}) are compatible for
  \begin{equation}\label{24cc}
\lambda=\frac{1}{2}.
\end{equation}
Also, it should be mentioned that $\beta=0$ results in
 \begin{equation}\label{24c}
\lambda=-\frac{3}{2}
\end{equation}
or
 \begin{equation}\label{24cd}
\lambda_1=0.
\end{equation}
In the following subsections, we explore these two categories of solutions.

\subsection{ Solutions with $\lambda=1/2 $ }\label{subsec1}

In this subsection, we study solutions related to $\lambda=1/2 $. Equations (\ref{18}) and (\ref{19})  admit the solutions
\begin{equation}\label{24d}
 \rho=\frac{-\alpha_1\pm\sqrt{\alpha^{2}_1+4\beta \alpha_2\frac{b'}{r^2} }}{2\beta\alpha_2}
\end{equation}
and
\begin{eqnarray}\label{25}
 \rho=\frac{-\alpha_3\pm\sqrt{\alpha^{2}_3+4\beta \alpha_4\frac{b}{r^3} }}{2\beta\alpha_4}.
\end{eqnarray}
Solutions (\ref{24d}) and (\ref{25}) are consistent while
 \begin{equation}\label{26}
\frac{\alpha_1}{\alpha_3}=\frac{\alpha_2}{\alpha_4}=m,
\end{equation}
and
\begin{equation}\label{27a}
\frac{b'}{b}=\frac{m}{r}.
\end{equation}
Equation (\ref{27a}) leads to
\begin{equation}\label{27}
b(r)=r^m
\end{equation}
which is one of the most famous shape functions in the wormhole theory \cite{phantom1}. This shape function provides all the necessary conditions to construct a traversable wormhole. This shape function has been introduced in \cite{Azizi, Moa, Shw, SR} using a linear EoS and $f(R, T)=R+\lambda T$, which satisfy the ECs. In reference \cite{squared} , the shape functions are selected freely; however, we assert that the only consistent solution is the power-law shape function, while other shape functions are inconsistent with the field equations. It is straightforward to show that, for $\lambda=1/2 $, combination of Eqs.(\ref{24b}) and (\ref{20a}) gives
\begin{equation}\label{2a8}
\alpha_1=1.8,\qquad\alpha_3=0.3-1.5\omega_r
\end{equation}
so
\begin{equation}\label{28}
m=\frac{6}{1-5\omega_r}.
\end{equation}
The asymptotically flat condition allows $m<1$ so Eq.(\ref{28}) leads to
\begin{equation}\label{28a}
\omega_r<-1,\qquad \omega_r>0.2.
\end{equation}

Now, we investigate solutions involving negative energy density. It is evident that the WEC and DEC are violated in this scenario, so we turn our attention to the NEC. One can conclude that for a negative energy density, $H>0$ holds true for
 \begin{equation}\label{1b}.
\omega_r<-1.
\end{equation}
On the other hand, $H_1>0$ holds for $\omega_t<-1$ which along with (\ref{24b}) show that
 \begin{equation}\label{2b}
\omega_r>1.4
\end{equation}
which is incompatible with condition (\ref{1b}). This result confirmed that solutions with negative energy density cannot fulfill the NEC and, consequently, other ECs.

Let us investigate the ECs for a positive  energy density that is more viable. From $H>0$, one can find
 \begin{equation}\label{2c}
\omega_r>-1.
\end{equation}
 It is easy to show that $H_1>0$ along with (\ref{24b}) give
\begin{equation}\label{1a}
\omega_r<1.4.
\end{equation}
Taking into account (\ref{28a}), (\ref{2c}) and (\ref{1a}) gives
\begin{equation}\label{2a}
0.2<\omega_r<1.4.
\end{equation}
 Considering (\ref{2a}) and (\ref{28}) results in
 \begin{equation}\label{2ab}
m<-1.
\end{equation}
One can use (\ref{24b}) and (\ref{2a}) to find
\begin{equation}\label{2ac}
-1<\omega_t<-0.4.
\end{equation}

\begin{figure}
\centering
  \includegraphics[width=3 in]{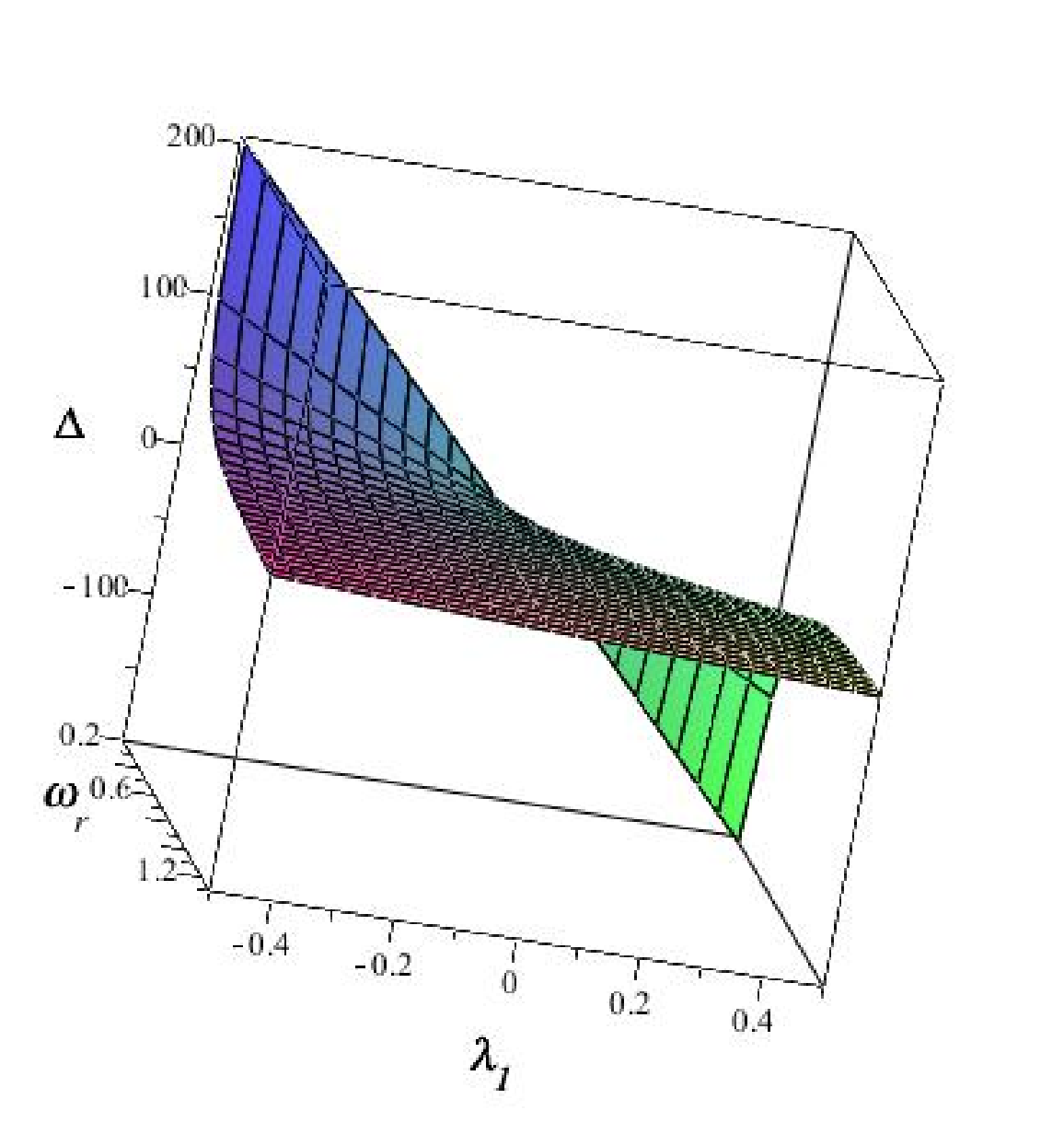}
\caption{The figure represents the $\Delta(r_0,\omega_r,\lambda_1)$ against $\omega_r$ and $\lambda_1$,  which is  negative in the entire range $0<\lambda_1$ and positive for $0>\lambda_1$ . See the text for details.}
 \label{fig1}
\end{figure}
\begin{figure}
\centering
  \includegraphics[width=3 in]{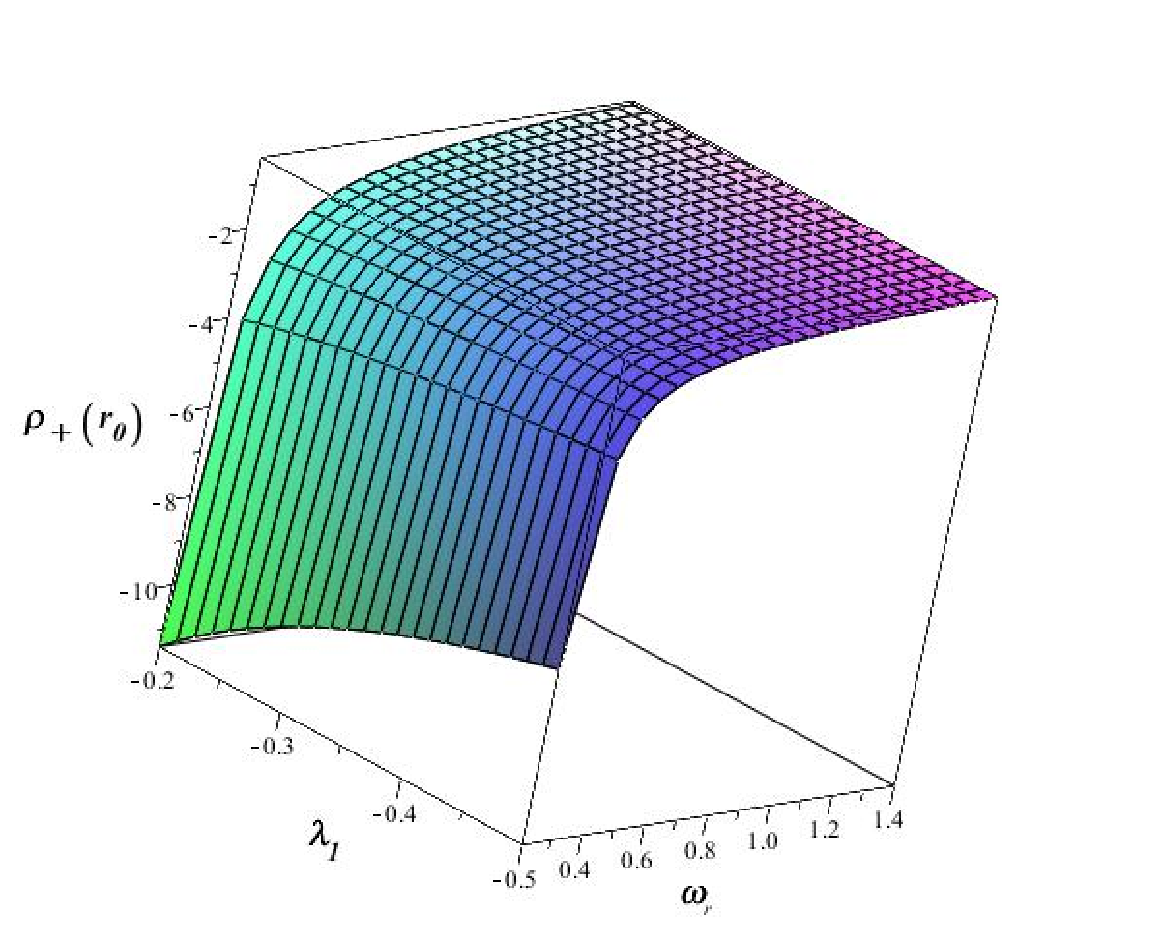}
\caption{The figure represents the $\rho_{+}(r=r_0)$ against $\omega_r$ and $\lambda_1$,  which is  negative in the entire range for $0.2<\omega_r<1.4$ and $\lambda_1<0$ . See the text for details.}
 \label{fig2}
\end{figure}
\begin{figure}
\centering
  \includegraphics[width=3 in]{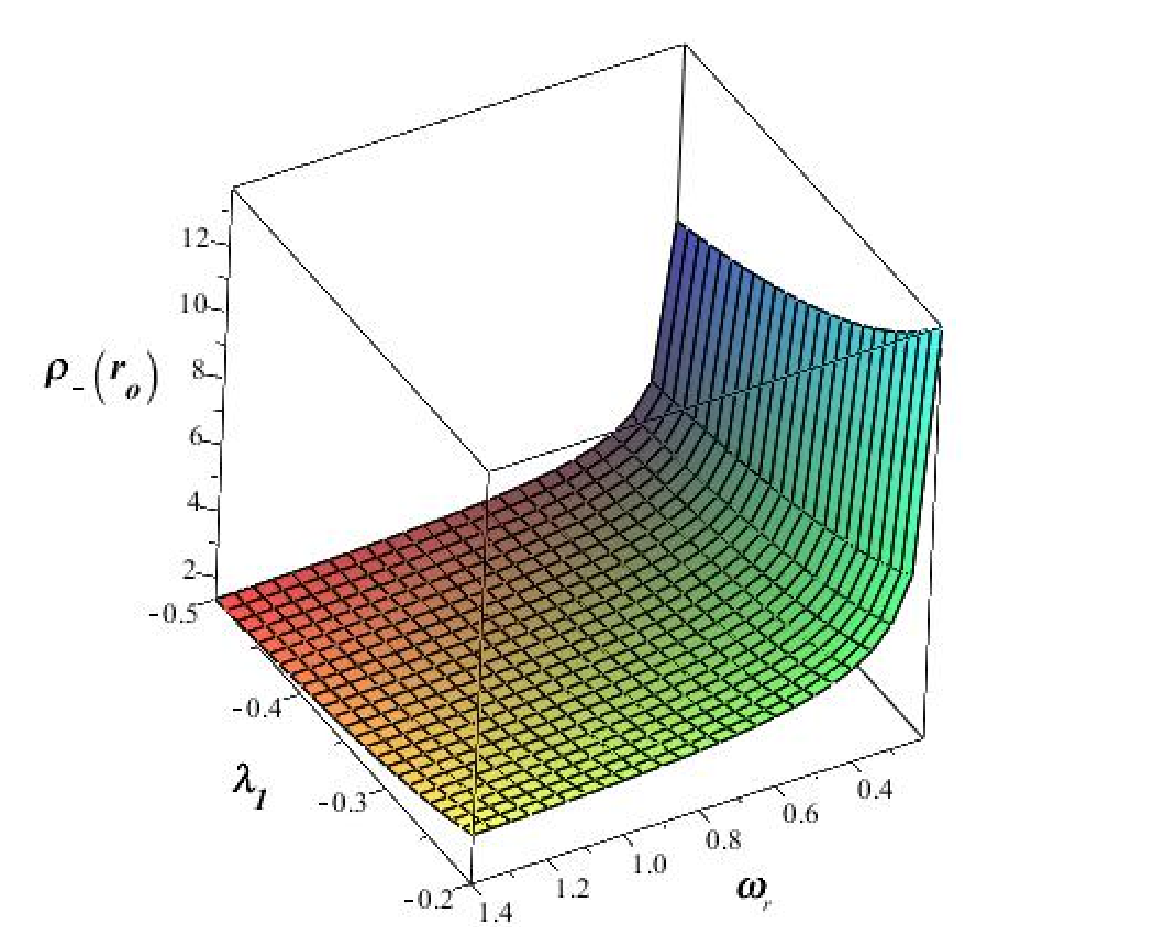}
\caption{The figure represents the $\rho_{-}(r=r_0)$ against $\omega_r$ and $\lambda_1$,  which is  positive in the entire range for $0.2<\omega_r<1.4$ and $\lambda_1<0$ . See the text for details.}
 \label{fig3}
\end{figure}
\begin{figure}
\centering
  \includegraphics[width= 3 in]{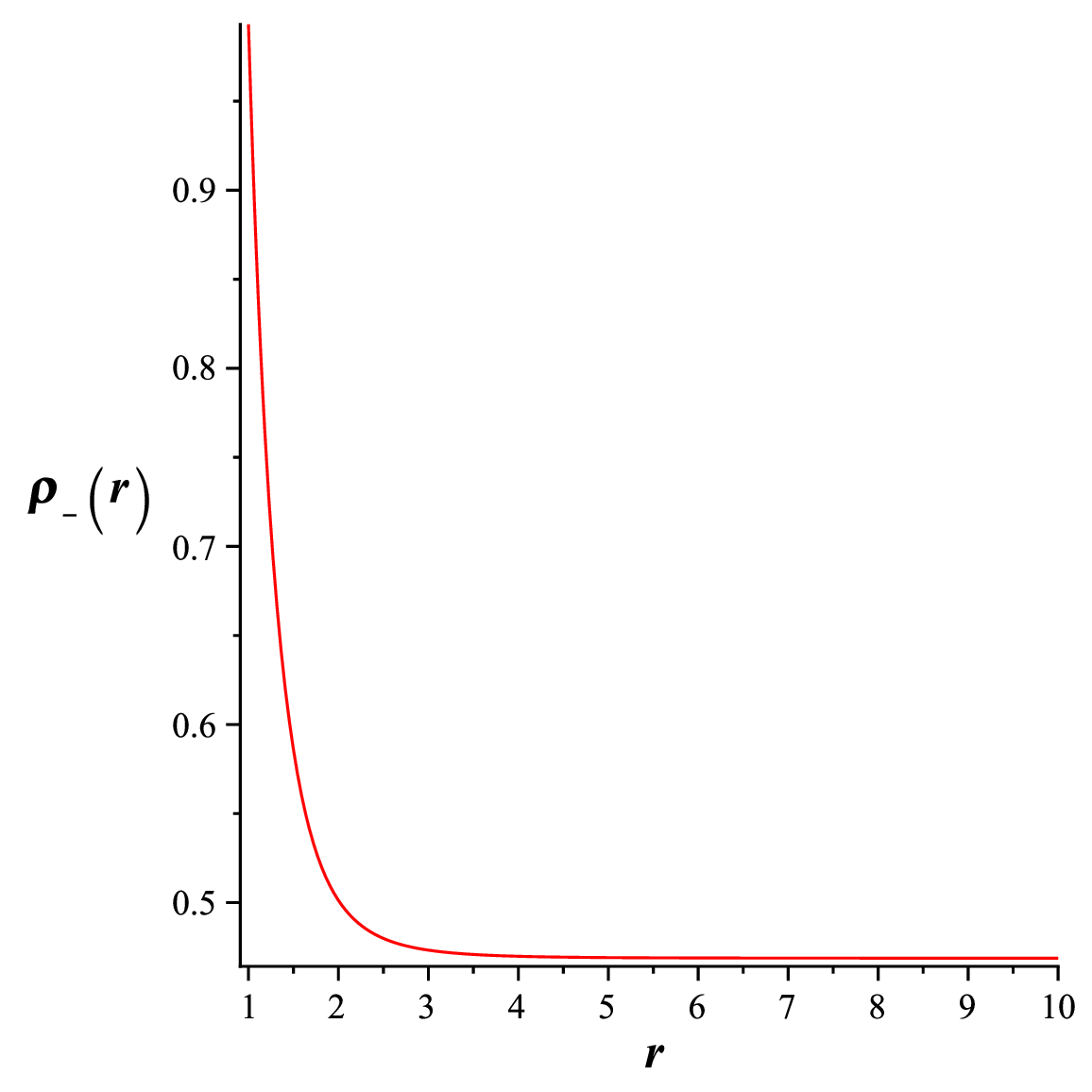}
\caption{The plot depicts $\rho_-(r)$ against radial coordinate for $m=-2$ and $\lambda_1=-1$,  showing a maximum at the throat and a monotonic decrease that approaches zero at large distances. See the text for details.}
 \label{fig4}
\end{figure}

 It is important to note that condition (\ref{2a}) and positive energy density fulfill the criteria necessary to satisfy all of the ECs. Now, let us investigate the energy density. Using (\ref{24b}) and (\ref{20a}) gives
\begin{equation}\label{3aa}
\alpha_1=1.8, \quad \alpha_2=14.4,\quad \beta=\frac{1.6}{6}\lambda_1
\end{equation}
 Equations (\ref{24d}), (\ref{28}) and (\ref{3aa}) result in
\begin{equation}\label{3a}
\rho_{\pm}=\frac{-1.8\pm\sqrt{\Delta(r,\omega_r,\lambda_1)}}{7.68\lambda_1}
\end{equation}
where
\begin{equation}\label{4a}
\Delta(r,\omega_r,\lambda_1)=1.8^2+\frac{92.16\lambda_1}{1-5\omega_r}r^{m-3}.
\end{equation}
We have ploted $\Delta$ as a function of $\omega_r$ and $\lambda_1$ at the wormhole throat in Fig.(\ref{fig1}). This figure demonstrates that $\Delta\geq0$ in the interval $0.2<\omega_r<1.4$ is reachable for
\begin{equation}\label{28b}
\lambda_1<0.
\end{equation}
Now, we can check the two possible energy densities at the wormhole throat for the + and - sign. We have plotted  $\rho_{+}(r_0)$ and  $\rho_{-}(r_0)$ against $\omega_r$ and $\lambda_1$  in Figs.(\ref{fig2}) and (\ref{fig3}). These figures show that $\rho_{-}(r_0)$  is positive, making it a suitable choice. Figure (\ref{fig3}) indicates that $\rho_{-}(r_0)$ increases as $\lambda_1$ increases and descends as $\omega_r$ decreases. As an example, the general behaviour of $\rho_{-}(r)$ against radial coordinate for $m=-2$ and $\lambda_1=-1$  is depicted in Fig.(\ref{fig4}). This figure illustrates that energy density is a monotonically decreasing function with a maximum at the wormhole throat, which approaches zero as $r\rightarrow \infty$. It is evident from Eqs.(\ref{2a}) and (\ref{2ac}) that tangential pressure is negative, whereas radial pressure is positive.
Wormholes with isotropic pressure   ($p_r=p_t$) are inaccessible in GR. It is straightforward to demonstrate that isotropic pressure is applicable for $\omega_r=-0.2$, which violates the asymptotically flat condition (\ref{28a}).

\begin{figure}
\centering
  \includegraphics[width=3 in]{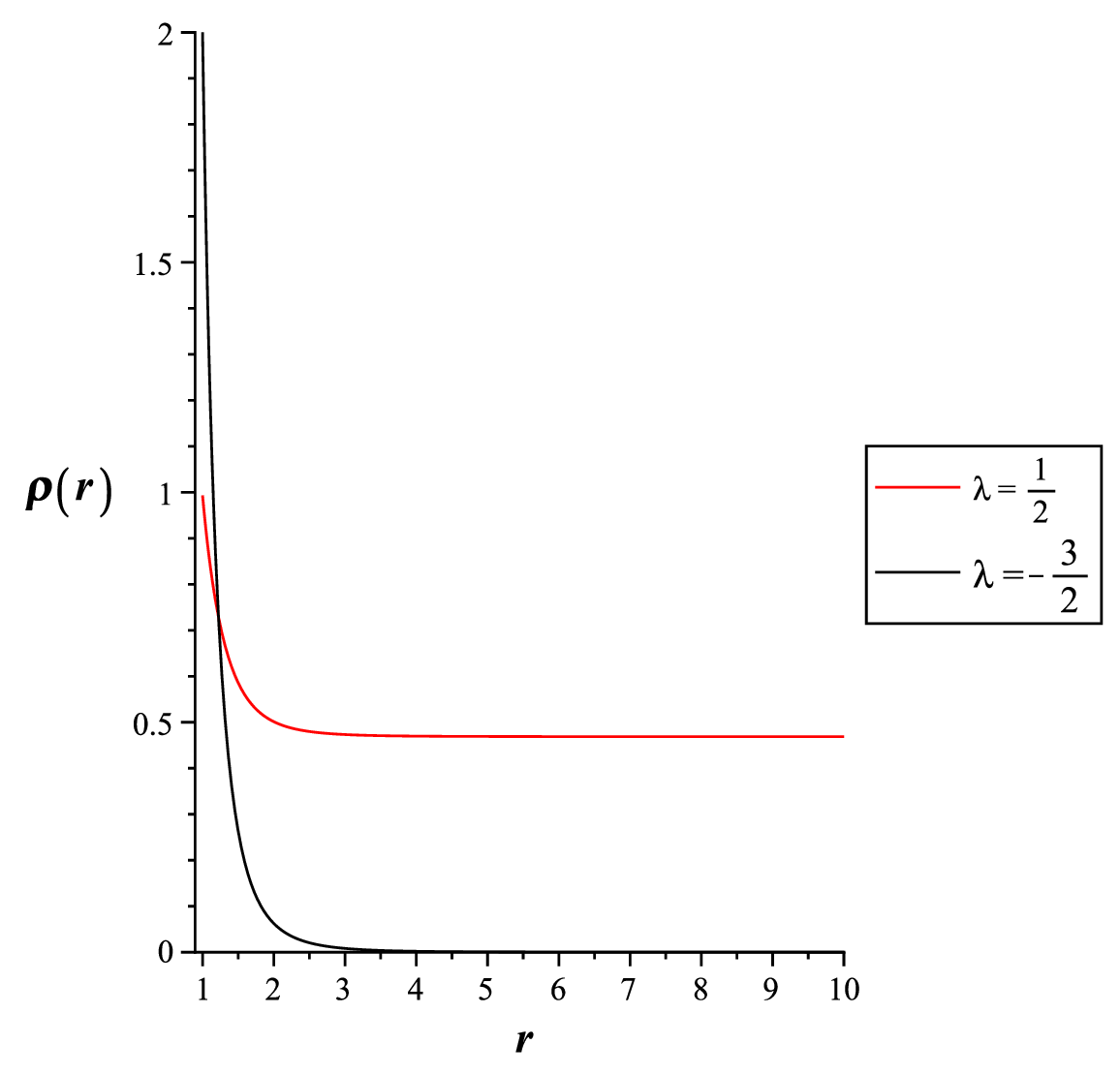}
\caption{$\rho(r)$ for $\lambda = 1/2$ (red), $\lambda = -3/2$ for $\lambda_ = -1$   (black) against $r$. It is clear that red line tends to a positive constant at large distances  while black line tends to zero. See the text for details}
 \label{fig5}
\end{figure}

\subsection{ Solutions with $\lambda=-3/2 $ }\label{subsec2}
Choosing $\lambda=-3/2$ leads to
\begin{equation}\label{28bb}
\omega_r=1-2\omega_t.
\end{equation}
In this case $\alpha_1$ and $\alpha_3$ are as follow
\begin{equation}\label{29}
\alpha_1=-1, \qquad\alpha_3=\frac{\omega_r-1}{2}.
\end{equation}
On the other hand, Eqs.(\ref{18}) and (\ref{19}) give
\begin{equation}\label{8b1}
\rho(r)=\frac{b'}{\alpha_1 r^2}
\end{equation}
and
\begin{equation}\label{30}
\rho(r)=\frac{b}{\alpha_3 r^3}.
\end{equation}
Using (\ref{8b1}) and (\ref{30}) results in
\begin{equation}\label{31}
\frac{b'}{b}=\frac{m}{r}
\end{equation}
where
\begin{equation}\label{32}
m=\frac{\alpha_1}{\alpha_3}=\frac{2}{1-\omega_r}.
\end{equation}
As in the previous case, the solution for (\ref{31}) is $b(r)=r^m$ but $m$ in (\ref{32})  defers from $m$ in (\ref{28}). Asymptotically flat condition yields
\begin{equation}\label{33}
\omega_r<-1, \qquad \omega_r>1.
\end{equation}
Let us investigate the NEC for negative energy density. In this case, $H>0$ shows that negative energy density is valid for
\begin{equation}\label{34}
\omega_r<-1
\end{equation}
on the other hand, $H_1>0$ is valid for
\begin{equation}\label{34b}
\omega_r>3.
\end{equation}
Thus, we can conclude that solutions with negative energy density cannot satisfy the ECs.

In the next step, we examine solutions with positive energy density. In this case, $H>0$ is valid for
\begin{equation}\label{35}
\omega_r>-1.
\end{equation}

Using  $H_1>0$ and(\ref{28bb}) gives
\begin{equation}\label{37}
\omega_r<3.
\end{equation}
Combination of (\ref{33}),(\ref{35}) and (\ref{37}) indicates that NEC is valid in the region
\begin{equation}\label{38}
1<\omega_r<3.
\end{equation}
One can also show that
\begin{equation}\label{39}
m<-1.
\end{equation}
is accessible for $1<\omega_r<3$. As in the previous case, it is straightforward to demonstrate that isotropic pressure is applicable for $\omega_r=-1/3$, which violates the asymptotically flat condition (\ref{33}).

\begin{figure}
\centering
  \includegraphics[width=3 in]{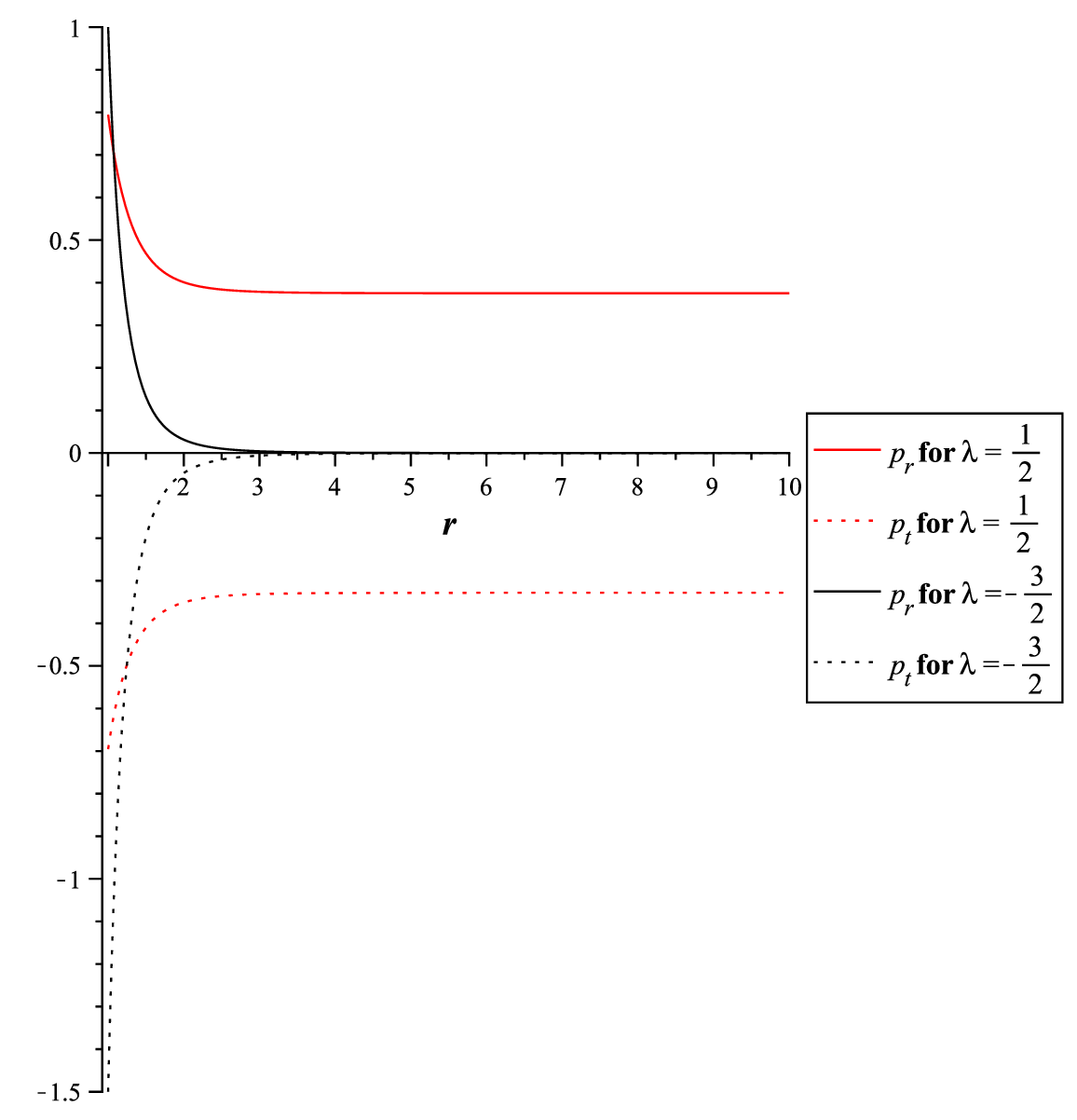}
\caption{$p_r(r)$ for $\lambda = 1/2$ (red solid line), $\lambda = -3/2$ for $\lambda_ = -1$   (black solid line), and $p_t(r)$ for $\lambda = 1/2$ (red dotted line), $\lambda = -3/2$ for $\lambda_ = -1$   (black dotted line) against $r$. It is clear that red line tends to a  constant at large distances  while black line tends to zero. See the text for details}
 \label{fig6}
\end{figure}

 Let us discuss the case $\beta=0$ in more detail. In this scenario, the solutions pertain to the cases $\lambda=-3/2$ or $\lambda_1=0$. It is clear that $\lambda_1=0$ simplifies the solutions to the linear form of $f(R,T)$ while the relation (\ref{24a}) holds. However, in the case of  $\lambda=-3/2$, the $f(R,T)$ becomes quadratic in the trace of the energy-momentum tensor, and the relation (\ref{24a}) is exclusive to$\lambda=-3/2$. A closer examination of this subject reveals that the case $\lambda_1=-3/2$ results in $T=0$. As an example, we have plotted $\rho$ as a function of $r$ for $m=-2$ for the cases $\lambda=1/2$ and  $\lambda=-3/2$ in Fig.(\ref{fig5}). Additionally, the $p_r$ and $p_t$ as functions of $r$ are depicted in Fig.(\ref{fig6}). These figures demonstrate that the same geometry can be sustained  with varying matter distributions in the $f(R,T)=R+\lambda T+\lambda_1 T^2$ scenario with different values for $\lambda_1$ and $\lambda$. It should be mentioned that energy density is a function of $\lambda_1$ for the case $\lambda=1/2$ while it is independent of $\lambda_1$ for the case $\lambda=-3/2$.

\section{Concluding remarks}

The wormhole has a significant drawback: the necessity for the wormhole spacetime to be traversable entails the flaring-out condition. Flaring-out conditions within the framework of GR result in a violation of the NEC, prompting the extensive use of modified theories to address the issue of exotic matter. Exotic matter, which challenges energy conditions, exhibits physical properties that would contradict established physical laws, such as a particle with negative mass. Recent research suggests that within modified gravity theories, it may be possible to construct wormholes using ordinary matter that adheres to all energy conditions. In recent years, significant progress has been made in modified theories of gravity, with researchers examining various extensions of GR. One prominent extension is $f(R,T)$ gravity, which modifies the traditional Einstein-Hilbert action by replacing the Ricci scalar with a function of the scalar curvature and the trace of the energy-momentum tensor. The literature has addressed wormholes within the framework of $f(R,T)$, where authors have investigated static spherically symmetric wormhole solutions for both linear and nonlinear models of $R$ and $T$.
 It has been demonstrated that $f(R,T)=R+\lambda T$ can be expressed as a spacetime with a negative Einstein’s gravitational constant, thereby eliminating the violation of ECs \cite{SR}. Subsequently, $f(R,T)=R+\lambda T$ demonstrates a similar formal framework to GR, but with a slight modification in Einstein's gravitational constant.

 In this work, we have examined $f(R,T)=R+\lambda T+\lambda_1 T^2$ to uncover new wormhole solutions. A linear dependence of $R$ in the $f(R,T)$ function provides the pure geometrical sector equivalent to GR. In the $f(R,T)$ modified theory of gravity, the right-hand side of Einstein’s field equations is altered due to the higher-order terms of $T$ in the $f(R,T)$ function. Consequently, only the material content sector is generalized. In $f(R,T)$ theory, the equations are expected to exhibit second-order characteristics in the metric coefficients, and the theory will remain free from the conventional instabilities that often plague numerous higher-order gravitational theories.
 Despite the modified field equations being quadratic in the matter quantities $\rho, p_r$, and $p_t$, the theory allows us to consider an EoS in a linear form.

  To achieve consistency in solutions, two possibilities exist:  $\beta\neq0$ and $\beta=0$. These two cases provide two categories of solutions.
   We have shown that $\omega_r$ and $\omega_t$ can be connected through Eq.(\ref{24a}). Authors in \cite{squared} claimed that an exponential shape function could be regarded as a solution but the power-law shape function has been demonstrated to be the sole possible solution in the context of squared trace extended gravity theory with vanishing redshift function. It has been demonstrated that solutions with negative energy density cannot fulfill ECs. To improve the viability of our solutions, we recommend a positive energy density.

  Solutions concerning $\beta\neq0$ are  valid solely for $\lambda=1/2$. In this class, solutions with positive energy density are valid for $\lambda_1<0$. It was shown that solutions with $m<-1$ can satisfy all ECs. The possible range for EoS parameters is found to be $0.2<\omega_r<1.4$ and $-1<\omega_t<-0.4$.  We have shown that energy density is a monotonically decreasing function with a maximum at the wormhole throat, which approaches zero as $r\rightarrow \infty$. Additionally, it was demonstrated that an increase in $\lambda_1$  can decrease the magnitude of energy density at the wormhole throat, while an increase in $\omega_r$ has the opposite effect.

   Solutions concerning $\beta=0$ can be divided into two parts: $\lambda_1=0$ and $\lambda=-3/2$. The case $\lambda_1=0$ is equivalent to $f(R,T)$ with a linear form of $T$. For the case, $\lambda=-3/2$, the $f(R,T)$ is a quadratic function of $T$. Similar to the case where $\lambda=1/2$, solutions with negative energy density cannot satisfy the ECs. In this case, the  possible range for EoS parameters are found to be $1<\omega_r<3$ and $-1<\omega_t<0$. Solutions are independent of $\lambda_1$. In this class of solutions, energy density is a monotonically decreasing function with a maximum at the wormhole throat, which approaches a non-vanishing constant  as $r\rightarrow \infty$.

   For bout classes of solutions,   it is evident that tangential pressure is negative, whereas radial pressure is positive. Also, asymptotically flat wormholes with isotropic pressure   ($p_r=p_t$) are inaccessible. In summary, it was shown the same geometry can be sustained  with varying matter distributions in the $f(R,T)=R+\lambda T+\lambda_1 T^2$ scenario with different values for $\lambda_1$ and $\lambda$.

We have demonstrated that  traversable wormhole solutions that satisfy all ECs exist within this theory, thus holding significant physical relevance. The approaches outlined in this study can be readily extended to accommodate more intricate dependencies of the function $f(R, T)$ in relation to $T$, provided that there are no crossed terms involving $R$ and $T$. In summary, the exploration of wormholes within the framework of $f(R,T)$ gravity represents a significant advancement in contemporary gravitational theories, potentially addressing enduring challenges in the fields of cosmology and astrophysics. In this context, we have analyzed a scenario with a vanishing redshift function, specifically $\phi(r)=0$, although it is also feasible to explore solutions that incorporate a non-constant redshift function.

\end{document}